\documentclass[conference]{IEEEtran}
\IEEEoverridecommandlockouts
\usepackage{cite}
\usepackage{amsmath,amssymb,amsfonts}
\usepackage{graphicx}
\usepackage{textcomp}
\usepackage{xcolor}
\usepackage{mathtools}
\usepackage{algorithm}
\usepackage{algpseudocode}
\algrenewcommand\algorithmicrequire{\textbf{Input:}}
\algrenewcommand\algorithmicensure{\textbf{Output:}}
\usepackage{geometry}
\def\BibTeX{{\rm B\kern-.05em{\sc i\kern-.025em b}\kern-.08em
    T\kern-.1667em\lower.7ex\hbox{E}\kern-.125emX}}

\renewcommand{\baselinestretch}{0.958}

\begin{document}

\title{Exploiting Beam Squint for Near-Field Wideband Beamforming in Multi-User OFDMA Systems
}

\author{\IEEEauthorblockN{Lanqing Zhi, Peilan Wang, and Jun Fang, \textit{Senior Member, IEEE}}
\thanks{This work was supported by the National Science and Technology Major Project of China under Grant 2025ZD1301900.}

\textit{University of Electronic Science and Technology of China}, Chengdu, China, 611731\\
Emails: lanqing\_zhi@std.uestc.edu.cn; peilan\_wangle@uestc.edu.cn; junfang@uestc.edu.cn;
}

\maketitle

\begin{abstract}
The beam squint effect has traditionally been viewed as detrimental to near-field wideband beamforming due to the reduction of beamforming gain towards the desired direction. In this paper, we show that it can instead be harnessed as a benefit in a hardware-constrained orthogonal frequency division multiple access (OFDMA) system, where the system, equipped with a single radio-frequency (RF) chain, aims to serve multiple spatially distributed users by allocating different subcarriers to different users. By exploiting the beam squint effect, flexible frequency-dependent beamfocusing patterns are realized via parallel true-time-delay (TTD) units, enabling nearly full beamforming gain across users. In particular, we formulate a max-min rate optimization problem that jointly designs the time delays, subcarrier-user association, analog beamforming, and power allocation. To tackle the resulting nonconvex mixed-integer problem, a two-stage optimization framework is proposed. In Stage~I, subcarrier-user association and structured analog beamforming are jointly optimized. In Stage~II, power allocation is obtained via bisection search and inverse water-filling. Simulation results demonstrate that the proposed method can achieve a significant improvement by utilizing beam squint to serve multiple spatially distributed users in OFDMA systems.  
\end{abstract}                       

\begin{IEEEkeywords}
Near-field communications, beam squint, true-time-delay, max-min fairness, resource allocation
\end{IEEEkeywords}

\section{Introduction}
With the development of millimeter-wave (mmWave) and terahertz (THz) communications, the abundant spectrum resources at high frequencies have provided strong support for ultra-high-rate wireless transmission \cite{zhang20196g}. However, high-frequency propagation suffers from severe path loss \cite{serghiou2022terahertz}. To compensate for this loss, extremely large antenna arrays (ELAAs) are deployed to form high-gain directional beams \cite{cui2022near}. 
As the array aperture increases, users are more likely to be located in spherical-wave-based near-field regions, where array response vectors depend on both angle and distance~\cite{cui2022channel,zhou2015spherical}.
In addition, the large bandwidth induces a non-negligible beam squint effect, which reduces the beamforming gain away from the boresight direction and consequently degrades performance, particularly in multi-user scenarios~\cite{cui2022channel,zhang2022beam}.

To address the beam squint issue in wideband systems, frequency-dependent delay compensation has been widely adopted together with different hybrid beamforming architectures. Specifically, the full-true-time-delay (Full-TTD) architecture and delay-phase precoding (DPP) architecture were proposed in \cite{monemi2025beam} and \cite{dai2022delay}, respectively, to mitigate beam deviation in conventional phase-shifter (PS)-based architectures. A sub-connected TTD-based architecture was proposed in \cite{wang2024beamfocusing} to reduce TTD and hardware requirements while improving energy efficiency. In \cite{ma2024switch}, switching-network-based designs such as the switch-based hybrid beamforming (SW-HBF) were investigated to enhance robustness against beam squint with lower power consumption and implementation complexity.

A plethora of prior works \cite{monemi2025beam,dai2022delay,wang2024beamfocusing,ma2024switch} have regarded beam squint as an undesirable effect and attempted to mitigate it. More recently, a line of research \cite{zhai2020thzprism,zhai2021ss} has reexamined beam squint in orthogonal frequency division multiple access (OFDMA) systems, showing that it can facilitate high-quality multi-user access with reduced hardware and algorithmic complexity. In particular, \cite{zhai2020thzprism} proposed a sequential true-time-delay (TTD) architecture, termed THzPrism, to generate frequency-dependent beam patterns that are pointing to different directions across different subcarriers, thereby extending angular coverage to serve multiple users. This approach was further extended to OFDMA systems with multiple radio-frequency (RF) chains \cite{zhai2021ss}, where multiple THzPrism beams are formed via hybrid beamforming to enable spatially spread multi-group transmission. 
Specifically, users are divided into multiple groups that share the same frequency band. While different groups are spatially separated to enable spectrum reuse, users within each group are served using frequency-dependent beam patterns. However, these approaches rely on a specific sequential TTD structure and typically result in a uniform frequency-dependent beam pattern. Consequently, they are primarily designed for far-field scenarios. In near-field conditions, where angle and distance are inherently coupled, such angle-based beam expansion cannot be readily generalized. Furthermore, existing studies predominantly emphasize coverage enhancement and improvements in system efficiency, whereas user fairness 
has received comparatively limited attention.


To address the above issues, this paper considers a wideband near-field multi-user downlink OFDMA system with a parallel TTD-based beamforming architecture. We aim to maximize the minimum rate among all users by jointly optimizing the TTD-based analog beamforming, subcarrier-user association, and power allocation. To solve the formulated nonconvex mixed-integer problem, we propose a two-stage optimization framework. In Stage~I, we jointly optimize the subcarrier-user association and analog beamforming under an equal-power surrogate. In Stage~II, we perform power allocation with the subcarrier-user association and beamforming fixed. Simulation results show that the proposed algorithm generates flexible frequency-dependent beamfocusing patterns that point to different spatial locations on different subcarriers, thus enabling an enhanced spatial diversity for multi-user data transmission.

\section{System Model and Problem Formulation}
\begin{figure}[!htbp]
  \centering
  \includegraphics[width=8cm,page=1,trim=0cm 0.8cm 0cm 0.8cm,clip]{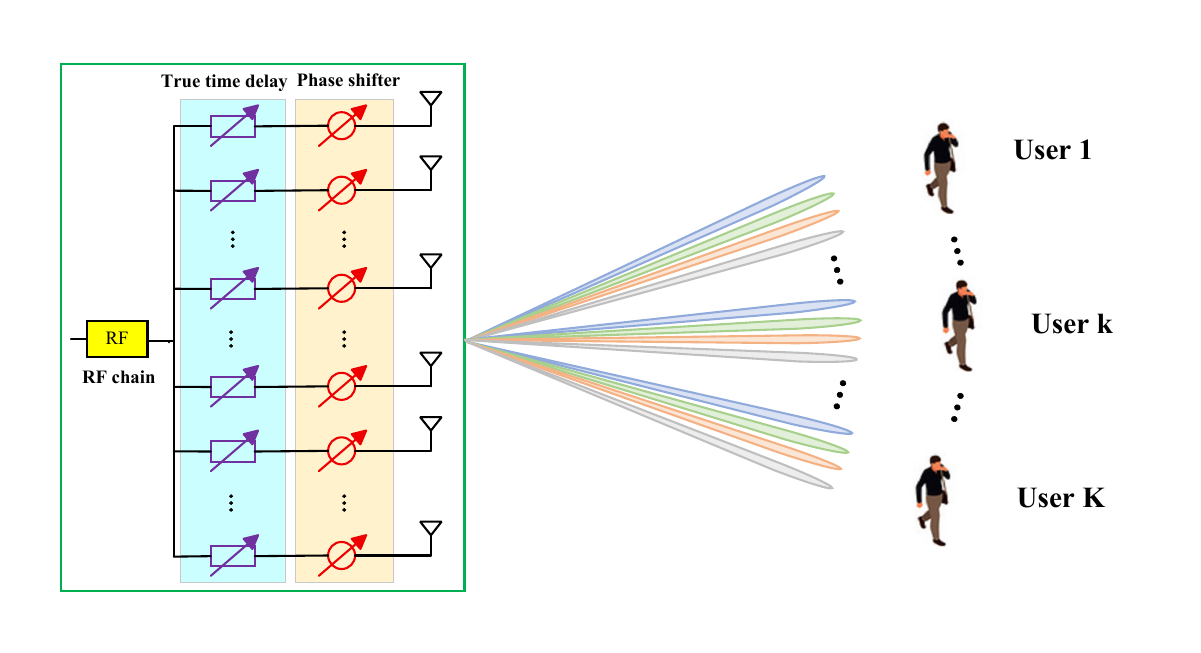}
  \caption{Parallel TTD-based beamforming structure.}
  \label{fig:structure}
\end{figure}

We consider a wideband near-field multi-user downlink OFDMA system, where a base station (BS), equipped with a single RF chain and an $N$-antenna uniform linear array (ULA), serves $K$ single-antenna users simultaneously. The central carrier frequency is denoted as $f_c$, and the wavelength is $\lambda_c=c_0/f_c$, where $c_0$ is the speed of light. The antenna spacing is set to $d=\lambda_c/2$. As shown in Fig.~\ref{fig:structure}, we consider a parallel TTD-based beamforming architecture, where each antenna is connected to both a phase shifter and a TTD unit to exploit the beam squint effect for multi-user access. The $n$-th TTD imposes a time delay $\tau_n \in [0, \tau_{\max}]$ on the signal, whereas the $n$-th phase shifter applies a phase shift $\phi_n$. Here, $\tau_{\max}$ denotes the maximum delay supported by each TTD unit. The total bandwidth $B$ is divided into $M$ subcarriers, where the frequency of the $m$-th subcarrier is $f_m = f_c - \frac{2m-1-M}{2M}B, m \in \{1,2,...,M\}$.

\subsection{Signal Model}
Owing to the property of the Fourier transform, the time delay $\tau_n$ introduced by the $n$-th TTD will result in a frequency-dependent phase shift $e^{-j2\pi f_m \tau_n}$ on the $m$-th subcarrier. Hence, the analog beamforming vector $\mathbf v_m \in \mathbb C^{N \times 1}$ at the $m$-th subcarrier can be written as
\begin{equation}
\mathbf{v}_m=\frac{1}{\sqrt N}\,\mathbf{w}\odot\mathbf{t}_m(\boldsymbol\tau),
\end{equation}
where $\mathbf{w}\triangleq[e^{j\phi_0},\ldots,e^{j\phi_{N-1}}]^T$ is the frequency-independent PS vector, with $w_n$ denoting the $n$-th element of $\mathbf{w}$, and $\mathbf{t}_m(\boldsymbol{\tau})\triangleq[e^{-j2\pi f_m\tau_0},\ldots,e^{-j2\pi f_m\tau_{N-1}}]^T$ is the frequency-dependent TTD vector. Here, $\boldsymbol{\tau}=[\tau_0,\ldots,\tau_{N-1}]^T$ denotes the TTD delay vector, and $\odot$ denotes the Hadamard product.

Let $\mathbf{s}_m=[s_{m,1},\ldots,s_{m,K}]^T$ denote the data symbol vector on the $m$-th subcarrier, where $s_{m,k}\sim\mathcal{CN}(0,1)$. Define $c_{m,k}\in\{0,1\}$ as the subcarrier-user association indicator, where $c_{m,k}=1$ means that subcarrier $m$ is assigned to user $k$, and $c_{m,k}=0$ otherwise. To eliminate intra-subcarrier multi-user interference, each subcarrier can be assigned to only one user, i.e.,
\begin{equation}
\sum_{k=1}^{K} c_{m,k} = 1,\quad \forall m.
\end{equation}

Then, the transmitted signal on the $m$-th subcarrier is given by
\begin{equation}
\mathbf{x}_m=\mathbf{v}_m\left(\sum_{k=1}^{K} c_{m,k}\sqrt{p_{m,k}}\,s_{m,k}\right),
\end{equation}
where $p_{m,k}\ge 0$ denotes the transmit power allocated to user $k$ on subcarrier $m$.

Define $\mathbf{h}_{m,k}$ as the channel vector between the BS and the $k$-th user on the $m$-th subcarrier. The corresponding received signal can thus be expressed as
\begin{equation}
\begin{aligned}
y_{m,k}
&=
\mathbf{h}_{m,k}^H\mathbf{x}_m+n_{m,k}\\
&=
\mathbf{h}_{m,k}^H\mathbf{v}_m
\left(\sum_{i=1}^K c_{m,i}\sqrt{p_{m,i}}\,s_{m,i}\right)
+n_{m,k},
\end{aligned}
\end{equation}
where $n_{m,k}\sim\mathcal{CN}(0,\sigma_k^2)$ denotes the additive white Gaussian noise at user $k$, and $\sigma_k^2$ represents the noise power at the $k$-th user. 
Since each subcarrier is exclusively assigned to one user, there is no intra-subcarrier multi-user interference. Therefore, the received signal-to-noise ratio (SNR) of the $k$-th user on the $m$-th subcarrier is defined as
\begin{equation}
\gamma_{m,k}
=
\frac{
p_{m,k}\left|\mathbf h_{m,k}^{H}\mathbf v_m\right|^2
}{
\sigma_k^2
}.
\label{eq:snr}
\end{equation}

Then, the achievable rate of the $k$-th user is given by
\begin{equation}
\begin{aligned}
R_k
&=\sum_{m=1}^Mc_{m,k}\log_2\!\left(1+\gamma_{m,k}\right).
\label{eq:user_rate}
\end{aligned}
\end{equation}

\subsection{Channel Model}

In this paper, we adopt a uniform spherical wave near-field channel model that comprises a line-of-sight (LoS) path and several non-line-of-sight (NLoS) paths \cite{akdeniz2014millimeter}. In particular, the channel between the BS and the $k$-th user on the $m$-th subcarrier can be modeled as
\begin{equation}
\begin{aligned}
    \mathbf{h}_{m,k}
&=
\sqrt{\frac{K_R}{K_R+1}}
\,\rho_{k,0}e^{-j2\pi f_m\delta_{k,0}}
\mathbf{a}(f_m,r_{k,0},\theta_{k,0}) \\
&+
\sqrt{\frac{1}{K_R+1}}
\sum_{\ell=1}^{L-1}
\rho_{k,\ell}e^{-j2\pi f_m\delta_{k,\ell}}
\mathbf{a}(f_m,r_{k,\ell},\theta_{k,\ell}),
\end{aligned}
\end{equation}
where $K_R$ denotes the Rician factor, $L$ is the number of paths, $\rho_{k,\ell}$ and $\delta_{k,\ell}
$ denote the complex gain and delay of the $\ell$-th path, respectively, $r_{k,\ell}$ and $\theta_{k,\ell}$ denote the distance and direction of the $\ell$-th scatterer, respectively. The index $\ell=0$ corresponds to the LoS path.

The near-field array response vector $\mathbf{a}(f,r,\theta)$ at $(r,\theta)$ can be expressed as
\begin{equation}
\mathbf{a}(f,r,\theta)
=
\frac{1}{\sqrt N}
\left[1,\phantom{0}
\cdots,\phantom{0} e^{-j\frac{2\pi f}{c_0}(r_{N-1}-r)}
\right]^T,
\end{equation}
where $r_n$ denotes the distance between the user and the $n$-th antenna element.

For a user located at polar coordinates $(r,\theta)$, the distance between the user and the $n$-th antenna is given by
\begin{equation}
r_n=\sqrt{r^2+n^2d^2-2rnd\sin\theta},\quad \forall n = 0,\cdots,N-1.
\end{equation}
Under the second-order Taylor approximation, we have
\begin{equation}
r_n \approx r+n^2d^2\alpha-nd\beta,\quad \forall n,
\end{equation}
where
\begin{equation}
\alpha=\frac{\cos^2\theta}{2r},\quad \beta=\sin\theta.
\label{par}
\end{equation}
Accordingly, the array response vector can be approximated as
\begin{equation}
\begin{aligned}
\mathbf{a}(f,r,\theta)
&\approx \mathbf{b}(f,\alpha,\beta) \\
&= \frac{1}{\sqrt N}
\Big[
1,\cdots,
e^{-j\frac{2\pi f}{c_0}\big((N-1)^2d^2\alpha-(N-1)d\beta\big)}
\Big]^T .
\end{aligned}
\end{equation}

\subsection{Problem Formulation}

In this work, our objective is to maximize the minimum achievable rate among all users by jointly optimizing the TTD-based analog beamforming, subcarrier-user association, and power allocation. The resulting optimization problem is formulated as
\begin{align}
\max_{\{c_{m,k}\},\{p_{m,k}\},\mathbf w,\boldsymbol\tau}
& \quad \min_{k} \quad R_k
\label{eq:main_problem} \\
\text{s.t.}\quad
& \mathbf v_m=\frac{1}{\sqrt N} \mathbf w\odot \mathbf t_m(\boldsymbol\tau),\quad \forall m,
\tag{\ref{eq:main_problem}a}\label{eq:main_problem_v} \\
& 0\le \tau_n\le \tau_{\max},\quad \forall n,
\tag{\ref{eq:main_problem}b}\label{eq:main_problem_tau} \\
& |w_n|=1,\quad \forall n,
\tag{\ref{eq:main_problem}c}\label{eq:main_problem_w} \\
& \sum_{k=1}^K c_{m,k}=1,\quad \forall m,
\tag{\ref{eq:main_problem}d}\label{eq:main_problem_csum} \\
& c_{m,k}\in\{0,1\},\quad \forall m,\,k,
\tag{\ref{eq:main_problem}e}\label{eq:main_problem_binary} \\
& p_{m,k}\ge 0,\quad \forall m,\,k,
\tag{\ref{eq:main_problem}f}\label{eq:main_problem_power_nonnegative} \\
& \sum_{m=1}^M\sum_{k=1}^K c_{m,k} p_{m,k}\le P_{\max},
\tag{\ref{eq:main_problem}g}\label{eq:main_problem_power_budget}
\end{align}
where $P_{\max}$ denotes the maximum transmit power budget.

Owing to the binary subcarrier-user association variables $\{c_{m,k}\}$ and the coupled analog beamforming variables $\mathbf{w}$ and $\boldsymbol{\tau}$, the above problem is a mixed-integer nonconvex optimization problem and thus difficult to solve directly. 
In the following, we develop a two-stage optimization framework. In Stage~I, we jointly optimize the subcarrier-user association and the analog beamforming under an equal-power surrogate. In Stage~II, with the subcarrier-user association and structured beamformers fixed, we perform a standard max-min power allocation update.

\section{Stage I: Joint Subcarrier-User Association and Analog Beamforming}

In this stage, we first initialize a feasible beamformer and then alternately optimize the subcarrier-user association and analog beamforming until convergence.

\subsection{Initialization of TTD and PS Parameters}

Since the channel state information (CSI) of all users is available, we can initialize the beamformer parameters to generate initial beams covering the target user region~\cite{zheng2024near}.
Based on the near-field parameters in~\eqref{par},
the target user region
$\theta\in[\theta_{\min},\theta_{\max}]$ and $r\in[r_{\min},r_{\max}]$
can be characterized in the $(\alpha,\beta)$ domain by
\begin{equation}
\alpha_{\min}
=
\frac{\min_{\theta}\cos^2\theta}{2r_{\max}},
\quad
\alpha_{\max}
=
\frac{\max_{\theta}\cos^2\theta}{2r_{\min}},
\label{eq:init_alpha_range}
\end{equation}
\begin{equation}
\beta_{\min}=\sin\theta_{\min},
\quad
\beta_{\max}=\sin\theta_{\max}.
\label{eq:init_beta_range}
\end{equation}

Let $f_L=\min_{m} f_m$ and $f_H=\max_{m} f_m$ denote the lowest and highest subcarrier frequencies, respectively. The initialization parameters $(\beta_0,\alpha_0,\hat{\beta},\hat{\alpha})$ are selected such that the beams at the two band edges approximately match the boundaries of the target region, i.e.,
\begin{equation}
\hat{\alpha}+\frac{f_c}{f_L}\alpha_0=\alpha_{\min},\quad
\hat{\alpha}+\frac{f_c}{f_H}\alpha_0=\alpha_{\max}.
\label{eq:init_alpha_match}
\end{equation}
\begin{equation}
\hat{\beta}+\frac{f_c}{f_H}\beta_0=\beta_{\max},\quad
\hat{\beta}+\frac{f_c}{f_L}\beta_0=\beta_{\min}.
\label{eq:init_beta_match}
\end{equation}
Based on these parameters, the initial TTD delays are set as
\begin{equation}
\tau_n
=
\frac{n^2d^2\hat{\alpha}-nd\hat{\beta}}{c_0},
\quad \forall n,
\label{eq:init_tau}
\end{equation}
where the initialized delays are chosen within the feasible range
$[0,\tau_{\max}]$ to satisfy the TTD hardware constraint.

The initial PS vector $\mathbf w$ is generated from the phase of the array response vector associated with $(\alpha_0,\beta_0)$ at the carrier frequency $f_c$, i.e.,
\begin{equation}
\mathbf w
=
\exp\!\big(j\,\angle(\mathbf b(f_c,\alpha_0,\beta_0))\big).
\label{eq:init_w}
\end{equation}
Hence, the initial structured beamformer on subcarrier $m$ is
\begin{equation}
\mathbf v_m
=
\frac{1}{\sqrt N} \mathbf w\odot \mathbf{t}_m(\boldsymbol\tau),
\label{eq:init_v}
\end{equation}
where $\mathbf{t}_m(\boldsymbol{\tau})=
\left[e^{-j2\pi f_m\tau_0},\ldots,e^{-j2\pi f_m\tau_{N-1}}\right]^T
\in\mathbb{C}^{N\times 1}$.

\subsection{Subcarrier-User Association}

Given the current structured beamformers $\{\mathbf v_m\}_{m=1}^M$ in Stage~I, we update the subcarrier-user association matrix $\mathbf C$ under an equal-power surrogate. Specifically, by fixing $\{\mathbf v_m\}$ and setting the power on each assigned subcarrier as $P_{\max}/M$, the rate of the $k$-th user in \eqref{eq:user_rate} can be rewritten as
\begin{equation}
R_k
=\sum_{m=1}^Mc_{m,k}\log_2\!\left(1+\frac{P_{\max}\left|\mathbf h_{m,k}^{H}\mathbf v_m\right|^2}{M \sigma_k^2}\right).
\label{eq:stage1_rate}
\end{equation}
Define
$r_{m,k} = \log_2\!\bigl(1+P_{\max}|\mathbf h_{m,k}^{H}\mathbf v_m|^2/(M\sigma_k^2)\bigr)$.
Then, \eqref{eq:stage1_rate} can be equivalently written as
\begin{equation}
R_k=\sum_{m=1}^M c_{m,k}r_{m,k}.
\label{eq:surrogate_rate}
\end{equation}
Collecting all $r_{m,k}$ yields the rate metric matrix
\begin{equation}
\mathbf R \triangleq [r_{m,k}] \in \mathbb{R}^{M\times K}.
\end{equation}

With fixed $\{\mathbf v_m\}$ and under the equal-power surrogate, the subcarrier
allocation problem in \eqref{eq:main_problem} can be formulated as
\begin{align}
\max_{\{c_{m,k}\},\,\eta}\quad
& \eta
\label{eq:sub_alloc}\\
\text{s.t.}\quad
& R_k \ge \eta,\quad \forall k,
\tag{\ref{eq:sub_alloc}a}\label{eq:sub_alloc_rate} \\
&~\eqref{eq:main_problem_csum},~\eqref{eq:main_problem_binary},
\tag{\ref{eq:sub_alloc}b}
\end{align}
where $R_k$ is given by \eqref{eq:surrogate_rate}.
The objective is to maximize the minimum accumulated rate among all users while satisfying the exclusive subcarrier-user association constraint. Due to the binary constraint in \eqref{eq:main_problem_binary}, problem \eqref{eq:sub_alloc} is a combinatorial optimization problem.

To make \eqref{eq:sub_alloc} tractable, we relax the binary constraint $c_{m,k}\in\{0,1\}$ to the convex constraint $0\le c_{m,k}\le 1$. The relaxed problem is given by
\begin{align}
\max_{\{c_{m,k}\},\,\eta}\quad 
& \eta \label{eq:sub_alloc_relaxed}  \\
\text{s.t.}\quad
& 0\le c_{m,k}\le 1,\quad \forall m,\, k,
\tag{\ref{eq:sub_alloc_relaxed}a}\label{eq:sub_alloc_relaxed_box} \\
&~\eqref{eq:main_problem_csum},~\eqref{eq:sub_alloc_rate}.
\tag{\ref{eq:sub_alloc_relaxed}b}
\end{align}
Problem \eqref{eq:sub_alloc_relaxed} is convex and can be efficiently solved by CVX \cite{grant2008graph}. Let $\mathbf C_{\mathrm{rel}}$ denote its optimal solution. In general, $\mathbf C_{\mathrm{rel}}$ is not binary and thus cannot be directly used as the final subcarrier-user association matrix.

To recover a feasible binary assignment from $\mathbf C_{\mathrm{rel}}$, we define the refining metric matrix as
\begin{equation}
\mathbf U \triangleq \mathbf C_{\mathrm{rel}} \odot \mathbf R,
\label{eq:refining_metric}
\end{equation}
where $\odot$ denotes the Hadamard product. The entry $U_{m,k}$ quantifies the priority of assigning subcarrier $m$ to user $k$, by jointly taking into account the relaxed assignment weight and the corresponding achievable rate.

Starting from $\tilde{\mathbf U}:=\mathbf U$, $\mathbf C:=\mathbf 0_{M\times K}$, and the accumulated-utility vector $\mathbf q:=\mathbf 0_{K\times 1}$ with $q_k$ denoting the $k$-th element of $\mathbf{q}$, the refining procedure is carried out iteratively. In each iteration, the user with the smallest accumulated utility is first selected, i.e.,
\begin{equation}
    k'=\arg\min_{k} \quad q_k.
\end{equation}
If multiple users attain the same minimum value, one of them is selected at random. Then, among all unassigned subcarriers, the subcarrier corresponding to the largest entry in the $k'$-th column of $\tilde{\mathbf U}$ is selected, namely
\begin{equation}
    m'=\arg\max_{m} \quad \tilde U_{m,k'}.
\end{equation}

Subcarrier $m'$ is then assigned to user $k'$ by setting $c_{m',k'}=1$, and the accumulated utility of user $k'$ is updated as
\begin{equation}
    q_{k'}=q_{k'}+U_{m',k'}.
\end{equation}

After that, the $m'$-th row of $\tilde{\mathbf U}$ is set to zero so that subcarrier $m'$ will not be selected again in subsequent iterations. The above procedure is repeated until all subcarriers have been assigned.

The resulting binary matrix $\mathbf C$ satisfies~\eqref{eq:main_problem_csum}, yielding a feasible exclusive subcarrier-user association. By prioritizing the user with the smallest accumulated utility, the refining procedure approximately preserves the max-min fairness implied by the relaxed solution.

\subsection{Analog Beamforming Optimization}
With the subcarrier-user association matrix $\mathbf C$ fixed, each subcarrier $m$ is assigned to a unique user $\kappa(m)$ satisfying $c_{m,\kappa(m)}=1$. From \eqref{eq:user_rate}, the achievable rate of the $k$-th user under the fixed subcarrier-user association and the equal-power surrogate is obtained as
\begin{equation}
R_k
=
\sum_{m:\kappa(m)=k}
\log_2\!\left(
1+\frac{P_{\max}\left|\mathbf h_{m,\kappa(m)}^{H}\tilde{\mathbf v}_m\right|^2}{M\sigma_{\kappa(m)}^2}
\right),
\label{eq:bf_surrogate_rate}
\end{equation}
where $\tilde{\mathbf v}_m$ denotes the ideal phase-only beamformer on subcarrier $m$.

Since $\tilde{\mathbf v}_m$ only appears in the rate term associated with user $\kappa(m)$ and does not affect the rates of the other users, the analog beamforming optimization decouples across subcarriers. Therefore, for each subcarrier $m$, the original max-min problem in \eqref{eq:main_problem} is equivalent to solving
\begin{equation}
\begin{aligned}
\max_{\tilde{\mathbf v}_m}\quad
& \left|\mathbf h_{m,\kappa(m)}^{H}\tilde{\mathbf v}_m\right|^2 \\
\text{s.t.}\quad
& |\tilde v_{m,n}|=\frac{1}{\sqrt N},\quad \forall n,
\end{aligned}
\label{eq:ideal_phase_problem}
\end{equation}
where $\tilde v_{m,n}$ denotes the $n$-th entry of $\tilde{\mathbf v}_m$.

For each subcarrier $m$, \eqref{eq:ideal_phase_problem} reduces to a phase-alignment problem. Accordingly, an optimal solution is given by
\begin{equation}
\tilde{\mathbf v}_m
=\frac{1}{\sqrt N}
\exp\!\big(j\,\angle(\mathbf h_{m,\kappa(m)})\big),
\label{eq:ideal_phase_solution}
\end{equation}
where the operators $\angle(\cdot)$ and $\exp(\cdot)$ are applied element-wise. 

The resulting ideal phase-only beamformers generally do not satisfy the parallel TTD-based beamforming architecture. We therefore project them onto the feasible set of the hardware architecture by solving
\begin{subequations}\label{eq:projection_problem}
\begin{align}
\min_{\mathbf w,\,\boldsymbol\tau}\quad
& \sum_{m=1}^M
\left\|
\tilde{\mathbf v}_m-\frac{1}{\sqrt N}\mathbf w\odot\mathbf t_m(\boldsymbol\tau)
\right\|_2^2 \\
\text{s.t.} \quad
&~\eqref{eq:main_problem_tau},~\eqref{eq:main_problem_w}.
\end{align}
\end{subequations}

Expanding the objective and removing constant terms, minimizing the objective function in~\eqref{eq:projection_problem} is equivalent to
\begin{equation}
\max_{\mathbf w,\,\boldsymbol\tau}
\sum_{n=0}^{N-1}
\Re\!\left\{
w_n\,s_n(\tau_n)
\right\},
\label{eq:projection_equivalent}
\end{equation}
where $s_n(t)\triangleq
\sum_{m=1}^M
\tilde v_{m,n}^* e^{-j2\pi f_m t}$.

For fixed $\boldsymbol\tau$, the phase shifter on each antenna branch admits the closed-form update
\begin{equation}
w_n
=
\exp\!\left(
-j\,\angle\!\big(s_n(\tau_n)\big)
\right),
\quad \forall n=0,\ldots,N-1.
\label{eq:w_update}
\end{equation}

For fixed $\mathbf w$, the delay update for each antenna element $n$ reduces to a one-dimensional optimization problem over $[0,\tau_{\max}]$, i.e.,
\begin{equation}
\tau_n
=
\arg\max_{t\in\mathcal Q}
\Re\!\left\{
w_n\,s_n(t)
\right\},
\quad \forall n=0,\ldots,N-1,
\label{eq:tau_update}
\end{equation}
where $\mathcal Q=\{0,\tau_{\max}/(Q-1),2\tau_{\max}/(Q-1),\cdots,\tau_{\max}\}$ denotes the search grid with $Q$ grid points.

The above procedure of alternately optimizing $\mathbf{w}$ and $\boldsymbol{\tau}$ is repeated until convergence.

Finally, the structured beamformer on subcarrier $m$ is updated as
\begin{equation}
\mathbf v_m=\frac{1}{\sqrt N}\mathbf w\odot\mathbf t_m(\boldsymbol\tau),\quad \forall m.
\label{eq:v_update}
\end{equation}

\section{Stage II: Power Allocation}
After Stage~I converges, let $\mathbf C^\star=[c_{m,k}^\star]$ and $\{\mathbf v_m^\star\}_{m=1}^M$ denote the subcarrier-user association matrix and structured beamformers, respectively. With $\mathbf C^\star$ and $\{\mathbf v_m^\star\}_{m=1}^M$ fixed, Stage~II optimizes the power allocation variables $\{p_{m,k}\}$. Then, the achievable rate of user $k$ can be expressed as
\begin{equation}
R_k
=
\sum_{m=1}^M
c_{m,k}^\star\log_2\!\left(
1+\frac{p_{m,k}\left|\mathbf h_{m,k}^H\mathbf v_m^\star\right|^2}{\sigma_k^2}
\right).
\label{eq:stage2_rate}
\end{equation}

Accordingly, the remaining max-min power allocation problem with fixed $\mathbf C^\star$ and $\{\mathbf v_m^\star\}_{m=1}^M$ is formulated as
\begin{align}
\max_{\{p_{m,k}\},\,t}\quad 
& t \label{eq:final_power_problem} \\
\text{s.t.}\quad
& R_k\ge t,\quad \forall k,
\tag{\ref{eq:final_power_problem}a}\label{eq:final_power_problem_a} \\
& \sum_{m=1}^M\sum_{k=1}^K c_{m,k}^\star p_{m,k}\le P_{\max},
\tag{\ref{eq:final_power_problem}b}\label{eq:final_power_problem_b} \\
&~\eqref{eq:main_problem_power_nonnegative}.
\tag{\ref{eq:final_power_problem}c}\label{eq:final_power_problem_c}
\end{align}
Problem \eqref{eq:final_power_problem} is convex and can be efficiently solved by existing methods, such as bisection on the common rate together with inverse water-filling~\cite{8836645,tse2005fundamentals}. 

\section{Computational Complexity Analysis}

Based on the above derivations, the overall procedure of the proposed method is summarized in Algorithm~\ref{alg:overall_method}. Since Stage~II only involves a standard convex power allocation with much lower complexity, the dominant computational complexity comes from Stage~I, including subcarrier-user association and analog beamforming.

In each alternating iteration, computing the rate metric matrix $\mathbf R$ requires $\mathcal O(MKN)$. Solving the relaxed subcarrier-user association problem \eqref{eq:sub_alloc_relaxed} by a standard interior-point method requires $\mathcal O\!\left((MK)^{3.5}\log(1/\epsilon)\right)$, where $\epsilon$ denotes the solution accuracy. In addition, the refining step requires $\mathcal O(MK+M^2)$. For analog beamforming optimization, constructing $\{\tilde{\mathbf v}_m\}$ requires $\mathcal O(MN)$, while the projection step requires $\mathcal O(I_pNMQ)$, where $I_p$ is the number of projection iterations. Updating $\{\mathbf v_m\}$ after convergence requires $\mathcal O(MN)$. 

Therefore, the total complexity of Stage~I can be expressed as $\mathcal O\!\left(
I_{\mathrm{AO}}
\left[
(MK)^{3.5}\log(1/\epsilon)+MKN+M^2+I_{p}NMQ
\right]
\right)$,
where $I_{\mathrm{AO}}$ denotes the number of alternating iterations.

\begin{algorithm}[t]
\caption{Proposed Two-Stage Optimization Algorithm}
\label{alg:overall_method}
\begin{algorithmic}[1]
\Require $\theta_{\min},\theta_{\max},r_{\min},r_{\max},\{\mathbf h_{m,k}\},\{\sigma_k^2\},P_{\max},\tau_{\max}$
\Ensure $\mathbf C^\star,\mathbf w^\star,\boldsymbol\tau^\star,\{\mathbf v_m^\star\},t^\star,\{p_{m,k}^\star\}$

\Statex \textbf{Stage I:}
\State Compute $(\beta_0,\alpha_0,\hat{\beta},\hat{\alpha})$ from \eqref{eq:init_alpha_match} and \eqref{eq:init_beta_match}
\State Initialize $\boldsymbol\tau$, $\mathbf w$, and $\{\mathbf v_m\}$ by \eqref{eq:init_tau}--\eqref{eq:init_v}
\Repeat
    \State Update $\mathbf C$ according to \eqref{eq:sub_alloc_relaxed} and \eqref{eq:refining_metric}
    \State Update the ideal beamformers $\{\tilde{\mathbf v}_m\}$ by \eqref{eq:ideal_phase_solution}
    \State Update $\boldsymbol\tau$ and $\mathbf w$ by \eqref{eq:tau_update} and \eqref{eq:w_update}
    \State Update $\{\mathbf v_m\}$ by \eqref{eq:v_update}
\Until{the surrogate minimum user rate converges}
\State Set $\mathbf C^\star=\mathbf C$, $\mathbf w^\star=\mathbf w$, $\boldsymbol\tau^\star=\boldsymbol\tau$, and $\{\mathbf v_m^\star\}=\{\mathbf v_m\}$

\Statex \textbf{Stage II:}
\State Obtain $t^\star$ and $\{p_{m,k}^\star\}$ for \eqref{eq:final_power_problem} by bisection and inverse water-filling

\State \Return $\mathbf C^\star,\mathbf w^\star,\boldsymbol\tau^\star,\{\mathbf v_m^\star\},t^\star,\{p_{m,k}^\star\}$
\end{algorithmic}
\end{algorithm}

\section{Simulation Results}

In this section, we evaluate the proposed framework through numerical simulations. Unless otherwise specified, the BS employs a $64$-element ULA and serves $K=4$ users over $M=10$ subcarriers. The carrier frequency and bandwidth are set to $f_c=24$~GHz and $B=8$~GHz, respectively. The maximum true-time delay is set to $\tau_{\max}=1$~ns~\cite{wang2024beamfocusing}, and the total transmit power is normalized to $P_{\max}=1$. Each user experiences $L=3$ propagation paths with a Rician factor of $K_R=13$ dB~\cite{akdeniz2014millimeter}.

We consider a near-field setting where the users are distributed in the angular sector $[-7^\circ,7^\circ]$ and distance range $[10,20]$ m. With the array aperture $D=(N-1)d$, the Fresnel and Rayleigh distances are
$R_{\mathrm{Fresnel}}\approx 0.62\sqrt{{D^3}/{\lambda_c}}\approx 1.37~\text{m}$
and
$R_{\mathrm{Rayleigh}}={2D^2}/{\lambda_c}\approx 24.8~\text{m}$,
respectively. The SNR is defined as $P_{\max}/\sigma^2$ with $\sigma^2=\sigma_k^2, \forall k$.

The benchmark schemes include \emph{PS-only}, center-focused beamforming (\emph{CFB}), and \emph{THzPrism}~\cite{zhai2021ss}. In the \emph{PS-only} scheme, all TTDs are disabled by setting $\tau_n=0$ for all $n$, with the remaining design procedure unchanged. In the \emph{CFB} scheme, the beam is focused on the center of the user region in the angle-distance domain. The \emph{THzPrism} benchmark follows the scheme in \cite{zhai2021ss} and adopts a serial TTD-based beamforming structure, where the phase shifters determine the beam center and the TTDs introduce frequency-dependent angular spreading to cover the user sector.



\begin{figure}[!htp]
	\centering
	  \includegraphics[width=0.4\textwidth]{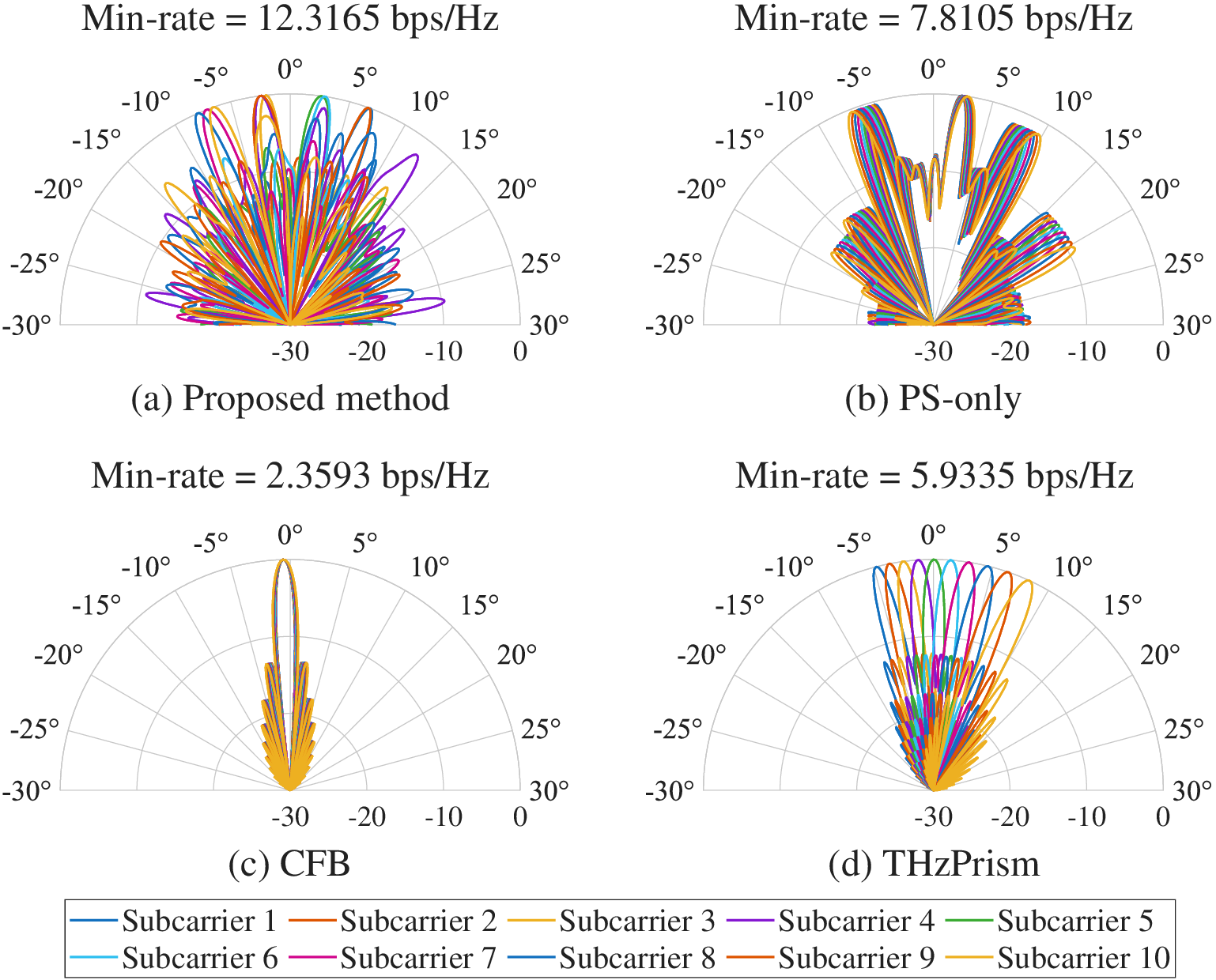}
	\hfil
    \caption{Polar beam patterns at $r=15$~m under $\mathrm{SNR}=30$~dB and $K=4$: (a) Proposed method, (b) PS-only, (c) CFB, (d) THzPrism.}
	\label{fig:beam_coverage}
\end{figure}

Fig.~\ref{fig:beam_coverage} shows the near-field polar beam patterns at $r=15$~m for a representative channel realization with $K=4$ users and $\mathrm{SNR}=30$~dB, where the user angles are uniformly distributed over $[-7^\circ,7^\circ]$. It is observed that the proposed method focuses beam energy more accurately toward the four user directions, thereby achieving the best user fairness, with the minimum rate among all users being 12.32 bps/Hz. In contrast, the \emph{PS-only} scheme produces multiple peaks with substantial overlap, while the \emph{CFB} scheme mainly concentrates on the center of the user region. Although \emph{THzPrism} can also form separated beams, it spreads the beams of different frequency components over the angular domain instead of truly assigning them to different users, resulting in poorer user focusing performance. Moreover, since it is based on far-field angular coverage and does not explicitly account for near-field distance variations, the resulting beam patterns may not accurately align with the actual user locations.

\begin{figure}[!htp]
	\centering
	\includegraphics[width=0.4\textwidth]{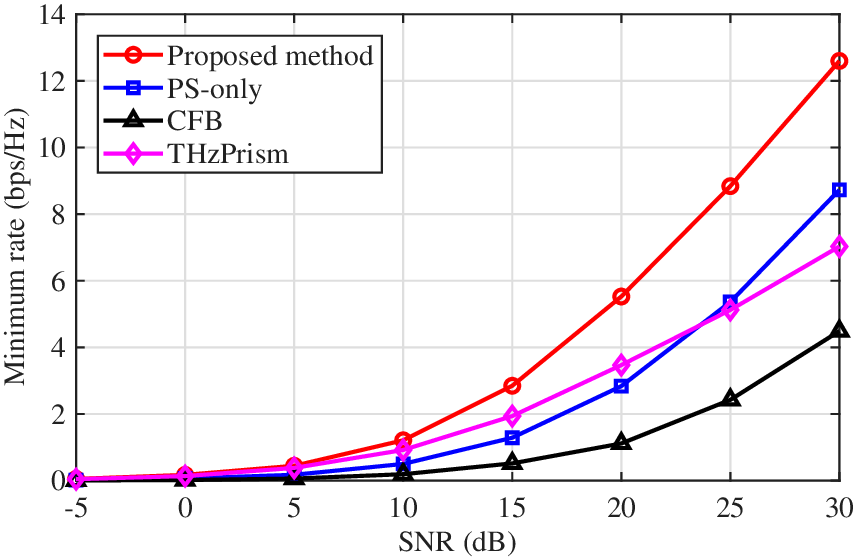}
	\hfil
    \caption{Minimum rate versus SNR when $K=4$.}
	\label{fig:SNR}
\end{figure}

Fig.~\ref{fig:SNR} presents the minimum rate versus SNR when $K=4$. As expected, the minimum rate of all schemes increases with SNR. Among them, the proposed method consistently achieves the highest minimum rate over the entire SNR range, and its advantage becomes more evident at high SNR. By contrast, \emph{CFB} 
performs the worst, while \emph{PS-only} and \emph{THzPrism} achieve intermediate performance.

\begin{figure}[!htp]
	\centering
	\includegraphics[width=0.4\textwidth]{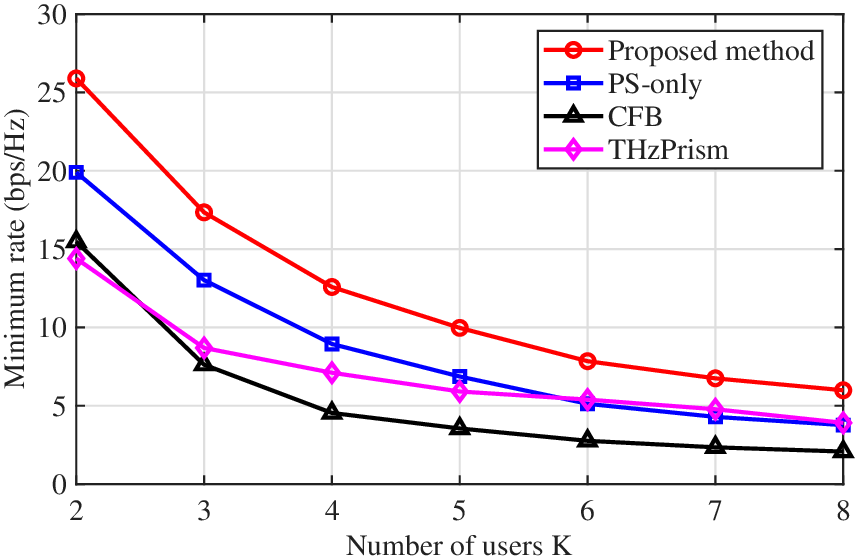}
	\hfil
    \caption{Minimum rate versus $K$ when $\mathrm{SNR}=30$ dB.}
	\label{fig:KK}
\end{figure}

As illustrated in Fig.~\ref{fig:KK}, the minimum rate is evaluated versus the number of users $K$ at $\mathrm{SNR}=30$~dB. It can be observed that the minimum rate decreases as $K$ increases, since the available resources must be shared among more users. Nevertheless, the proposed method maintains the highest minimum rate over the entire range of $K$, demonstrating its superior max-min fairness performance under increasing system load. In contrast, \emph{CFB} performs the worst in most cases, while \emph{PS-only} and \emph{THzPrism} remain inferior to the proposed method.

\section{Conclusions}
In this paper, we investigated fair beamforming for wideband near-field multi-user downlink OFDMA systems. Under a single-RF-chain architecture with parallel TTDs, a two-stage optimization framework was developed for joint subcarrier-user association, analog beamforming, and power allocation. Simulation results demonstrated that the proposed scheme achieves superior beam coverage and minimum rate, highlighting the potential of TTD-enabled delay-domain flexibility for fair transmission in wideband near-field OFDMA systems.

\bibliographystyle{ieeetr}
\bibliography{ref} 
\end{document}